# X-ray Reflectivity Studies of Atomic-level Surface-segregation in a Liquid Eutectic Alloy of AuSn

Venkatachalapathy S. K. Balagurusamy[1], Reinhard Streitel[1], Oleg G. Shpyrko[2], P. S. Pershan[1], Mati Meron[3] and Binhua Lin[3]

[1]Department of Physics and DEAS, Harvard University,
Cambridge, MA 02138, USA
[2]Center for Nanoscale Materials, Argonne National Laboratory,
Argonne, IL 60439
[3]CARS, University of Chicago,
Chicago, IL 60637, USA

Abstract

X-ray reflectivity studies reveal atomic-level surface-segregation at the free surface of the eutectic **$Au_{71}Sn_{29}$** liquid alloy. The surface-segregation extends up to three layers, in which the top layer is almost a pure monolayer of Sn, the second layer is almost a pure monolayer of Au and the third layer appears to be slightly enhanced in Au. Although the surface-segregation concentration profiles can be qualitatively accounted for by the theories of Defay-Prigogine and Strohl-King, they cannot satisfactorily account for the measured surface tension.

# 1. **INTRODUCTION**

The Gibbs adsorption rule predicts that for liquid mixtures of two elements with disparate surface tensions the surface concentration of the component with the lower surface tension will be enhanced. [1-3]. This rule has been successfully applied to explain the observed segregation of the low-surface tension component to the surface in numerous binary liquid mixtures[2]. Until recently most of the studies have examined the surface-segregation behavior by measuring either adsorption isotherms or surface tension values [2]. Unfortunately, these methods do not reveal the kind of details about the depth dependence of the atomic segregation near the surface that is necessary for full understanding of the interfacial electronic and thermodynamic properties. On a practical level alloys such as InSn, AuSn and BiSn are typical of the wide class of materials that are being actively studied as substitutes for toxic Pb-based solder in electronics applications and as packaging materials with good mechanical properties and there is need for more specific information on the nature of their surfaces [4,5]. Although some insights in the near surface regions of the liquid/solid interface can be obtained by transmission electron microscopy measurements of nano-size alloy particles such as those carried out by Mori and colleagues on just these alloys[6,7] these techniques can not provide the details of the atomic-level segregation near the surface layers of liquid alloys that can be obtained from X-ray reflectivity studies of the type to be presented here.[8-11].

For example, X-ray studies have shown that the GaBi liquid alloy that has a large positive enthalpy of mixing, 1.05kJ/mole[12] exhibits a rich surface behavior that includes surface-segregation of the lower surface tension Bi into a monolayer followed by a thick wetting film of a Ga-rich liquid between the Bi monolayer and the bulk alloy[8,9,13].

X-ray results from another alloy, BiSn[11] for which the positive enthalpy of mixing is smaller, 0.105kJ/mole[14] shows surface-segregation in the top three layers with the top most layer very rich in Bi, the next layer rich in Sn followed by a third Bi rich layer. In contrast the BiIn liquid alloy with negative enthalpy of mixing -1.212kJ/mole[14], shows only modest surface-segregation with the low surface tension Bi in the top monolayer[10]. We report here X-ray reflectivity studies of a AuSn liquid alloy that has large negative enthalpy of mixing (-9.66 kJ/mole)[14] and atomic-level surface segregation in the top three layers.

## 2. EXPERIMENTAL METHOD

Solid ingots of $Au_{71}Sn_{29}$ alloy sample (*Alfa Aesar, 99.99% purity)* were placed in a UHV chamber in a Mo pan whose surface was cleaned by sputtering with $Ar^+$ ions. The UHV chamber was evacuated to $10^{-9}$ Torr and the bakeout process was started. During the bakeout, the walls of the UHV chamber and its components were heated gradually to 100-200°C to desorb moisture. Following about 2 days of bakeout, the chamber was cooled to room temperature. A Boraelectric heating element mounted underneath the sample pan was used to heat the sample to temperatures of the order of 330°C to melt the sample. At this point the macroscopic native oxide, if any, as well as any possible contaminations at the surface were removed by mechanical scraping of the liquid surface with a Mo strip wiper. A clean liquid surface was finally obtained by further sputtering by $Ar^+$ ion beams for several hours to remove the remaining microscopic surface oxide.

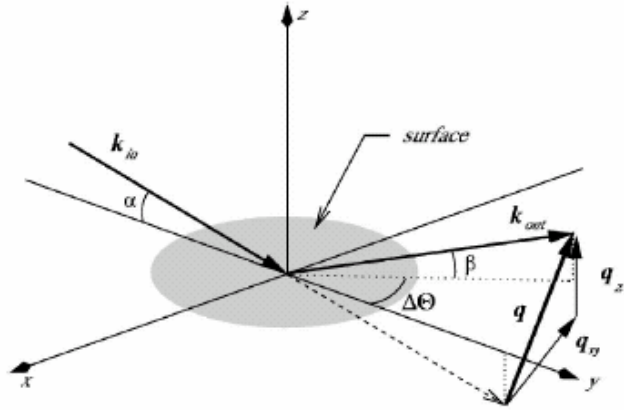

**Figure 1** Kinematics of the X-ray measurements from the sample. The incident and detected X-ray wave vectors are $k_{in}$ and $k_{out}$ respectively.

The X-ray reflectivity measurements from a clean liquid surface of the alloy were carried out at the liquid surface spectrometer facility at ChemMat CARS beamline at the Advanced Photon Source in Argonne National Laboratory. The kinematics of these X-ray measurements are illustrated in **Figure 1.** X-rays with a wavelength of $\lambda=0.943$Å corresponding to the wavevector $k_{in}=2\pi/\lambda$, impinge on the horizontal liquid surface at an incident angle $\alpha$. The detector captures the outgoing ray with a wavevector $\vec{k}_{out}(\beta,\Delta\Theta)$. The specular reflectivity $R(q_z)$ is the ratio of the intensity of the outgoing ray to the incident intensity, when the conditions for specular reflection, $\alpha=\beta$ and $\Delta\Theta=0$ are fulfilled. In this case the surface-normal momentum transfer is $q_z=(4\pi/\lambda)\sin\alpha$. The detector resolution due to a finite angle of acceptance was defined by horizontal and vertical slits mounted in front of the detector, approximately 650 mm from the sample. A second set of slits located approximately 200 mm from the sample on the detector arm was used to shield against unwanted parasitic scattering. During all specular reflectivity measurements the full widths of the angular acceptance of the detector were fixed at 9.4 mrad vertically and 3.1 mrad horizontally. The measurements were carried out with the sample held at a temperature of 295°C with a stability better than 0.05°C. Mechanical vibrations from the liquid surface spectrometer stage were isolated from the UHV sample chamber by an active feedback-controlled vibration isolation mechanism.[15] The reflected intensity was measured with an Oxford scintillation detector.

For low $q_z$ (<~0.6Å$^{-1}$) range, the specular reflectivity is obtained by normalizing the reflected signal to the incident beam intensity. The procedure at high $q_z$ (>~0.6Å$^{-1}$) is more complicated because of the appreciable background due to the isotropic scattering from the bulk of the liquid. This background has a broad peak centered at nearly the same $q_z$ as the surface-induced layering peak. Therefore, for $q_z \geq 0.6$Å$^{-1}$ a background-subtracted specular signal is represented by the difference between the signal measured at ($\alpha=\beta$ and $\Delta\Theta=0$) and the off-specular signal that is collected at $\Delta\Theta=\pm 0.3°$ (~5 mrad). For $q_z<0.6$Å$^{-1}$ the relative contribution of the bulk liquid scattering to specular reflection is sufficiently small that it can be neglected.

The fraction of the X-ray intensity reflected at an angle $\alpha$ from the liquid surface is given in the Born approximation by[16, 17]

Eq. 1     $R(q_z) = R_F(q_z) CW(q_z,T) |\Phi(q_z)|^2$

where $R_F(q_z)$ is the theoretical Fresnel X-ray Reflectivity from an abrupt flat interface between vacuum and the bulk alloy. The function $CW(q_z,T)$ is a Debye-Waller like factor due to thermally excited capillary waves. The procedure for calculating $CW(q_z,T)$ has been detailed in earlier papers on surface layering in Ga[18], In[16], K[17], and water[19]. The basic idea is to convolve the known resolution with the theoretical algebraic form

$\sim 1/\left|\vec{q}_{xy}\right|^{2-\eta}$ where

$$\left|\vec{q}_{xy}\right|^2 = (2\pi/\lambda)^2 \left[\cos^2\alpha + \cos^2\beta - 2\cos\alpha\cos\beta\cos2\Delta\Theta\right]$$

and $\eta = (k_B T/2\pi\gamma) q_z^2$ where for α≠β $q_z=(2\pi/\lambda)(\sin\alpha+\sin\beta)$. The actual form used in Eq. 1 is the difference between the convolution at α=β and ΔΘ=0 and the average of the convolution at the offset angles of ΔΘ=±0.3°.

The surface tension for the alloy can be determined from the η dependence of the diffuse scattering from thermal capillary waves.[17, 19] **Figure 2** contains a comparison between the background subtracted diffuse scattering data (filled circles) and the β dependence of a background-subtracted convolution of the theoretically derived scattering cross-section (solid line). The narrow detector vertical resolution of 1.55 mrad for α=5.07° applicable for both of these was typical of scans taken in order to measure the surface tension γ. Note that this resolution is considerably smaller than the vertical resolution of the detector, 9.5 mrad, that was used to measure R($q_z$). This data corresponds to $q_z$=1.1 Å$^{-1}$ and the solid line is calculated for the value of surface tension of 615 mN/m that gives the best fit to the data. Broken lines corresponding to γ=(600, 640) mN/m illustrate the confidence limits of fit. The other broken line (– . –) illustrates the calculation for the mean value of the published values of the surface tension, γ=750 mN/m.[20] The structure factor obtained from the measured reflectivity uses the value γ=615 mN/m.

The surface structure factor, Φ($q_z$) is given by the Fourier transform of the surface-normal derivative of the electron density $\rho(z)$,[21, 22]

Eq. 2    $$\Phi(q_z) = \frac{1}{\rho_\infty} \int dz \frac{d\langle \rho(z) \rangle}{dz} \exp(iq_z z)$$

where $\langle\rho(z)\rangle$ is the thermal average of the surface-normal electron density over the illuminated portion of the surface at a vertical position z and $\rho_\infty$ is the electron density in the bulk liquid. Since the form of $R_F(q_z)$ is determined completely by the critical angle for total external reflection, and $CW(q_z,T)$ is known accurately from capillary wave theory, the intrinsic electron density profile, $\langle\rho(z)\rangle$, can be obtained by numerically curve-fitting the measured $R(q_z)$ to a physically motivated model.[10, 11]

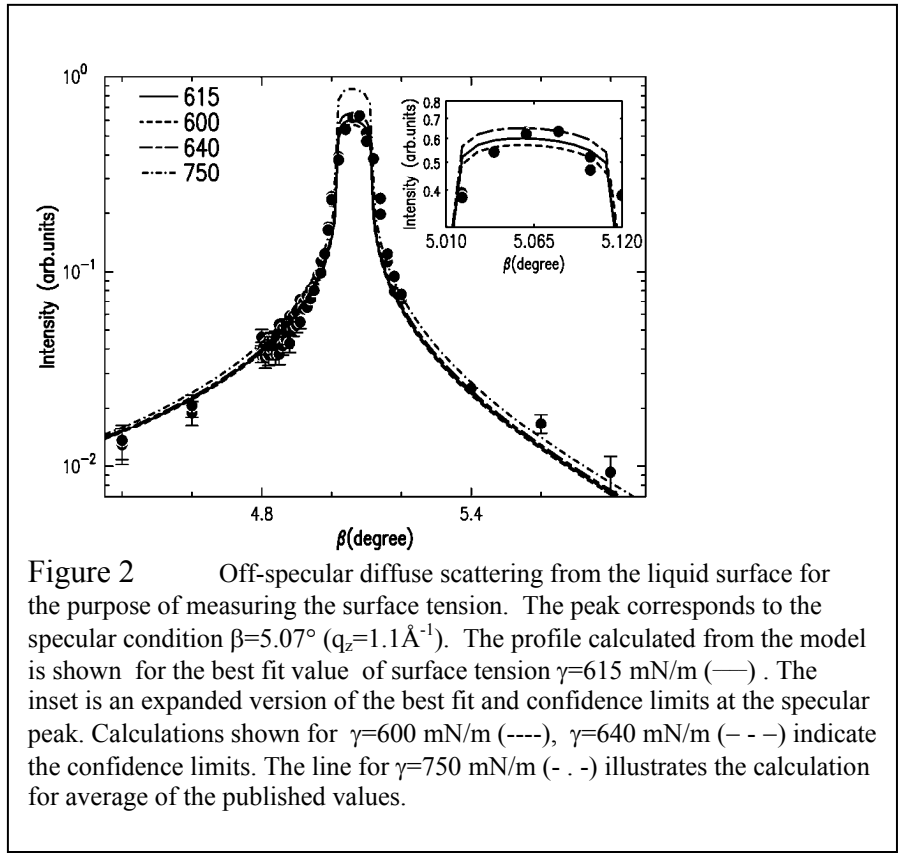

Figure 2    Off-specular diffuse scattering from the liquid surface for the purpose of measuring the surface tension. The peak corresponds to the specular condition β=5.07° ($q_z$=1.1Å$^{-1}$). The profile calculated from the model is shown for the best fit value of surface tension γ=615 mN/m (——) . The inset is an expanded version of the best fit and confidence limits at the specular peak. Calculations shown for γ=600 mN/m (----), γ=640 mN/m (– - –) indicate the confidence limits. The line for γ=750 mN/m (- . -) illustrates the calculation for average of the published values.

The distorted crystal (DC) model provides a very good description of the single surface-layering peak in X-ray reflectivity at $q_z \approx \pi/a$ (a: atomic radius) that is observed for many elemental liquid metals [16-18, 23]. The simplest form of the model for the electron density consists of a series of Gaussian peaks centered at $z_n = d_0(n-1)$ for $n=1,2,3,\ldots$. In the DC model the Gaussians have common integrated densities and widths that increase linearly with distance from the surface. For example, if the width of the Gaussian peak in the $n^{th}$ layer is given by $\sigma_n^2 = \sigma_0^2 + (n-1)\bar{\sigma}^2$ these Gaussian peaks become broader and merge to represent the bulk density at large n. This model naturally describes the decay of surface induced layering with distance from the surface. Furthermore, this model has the advantage that the Fourier transform of Eq. 2 can be computed analytically. The Fourier transform of this series of equally-spaced Gaussian peaks gives rise to the surface-layering peak that is observed in the reflectivity. Additional features, other than a simple surface-layering peak, that are observed in binary liquid metal alloys Bi-In[10] and Bi-Sn[11] can be described by modifying either the integrated densities in individual layers or the relative position of the top few layers.

The modified electron density model that is used here to describe the observed dependence on $q_z$ of the X-ray reflectivity from the $Au_{71}Sn_{29}$ liquid alloy surface differs from the standard DC model in that it treats amplitudes, positions and roughness for the first three layers as independent parameters. The form for the modified model is

**Eq. 3** $$\frac{\langle \rho(z) \rangle}{\rho_\infty} = \sum_{n=1}^{\infty} \frac{e^{-(z-z_n)^2/\sigma_n^2}}{\sqrt{2\pi}\sigma_n / d_0} \left( (1-x_n)\rho_{Au}^{rel} + x_n \rho_{Sn}^{rel} \right) \frac{C_n}{C}$$

where the positions of the first three layers, $z_1, z_2$ and $z_3$, and their widths, $\sigma_1$, $\sigma_2$, and $\sigma_3$ are adjustable parameters. For $n \geq 4$ the layer positions and widths of the Gaussians are $z_n = z_3 + (n-3)d_0$ and $\sigma_n^2 = \sigma_0^2 + (n-1)\bar{\sigma}^2$. The amplitudes of the Gaussians are expressed in terms of effective X-ray energy dependent electron densities of Au and Sn atoms, $\rho_{Au}^{rel}, \rho_{Sn}^{rel}$, relative to the bulk alloy. They are obtained from the real part of the atomic scattering factors, Z'(E).[10, 11] The concentration of Sn atoms in the $n^{th}$ layer is $x_n$ and $C_n \left( x_n v_{Sn} + (1-x_n) v_{Au} \right) = 1$ for the $n^{th}$ layer and $C \left( x_\infty v_{Sn} + (1-x_\infty) v_{Au} \right) = 1$ for the bulk, where C and $C_n$ are the average atomic densities in the bulk and n-th layers respectively, and $v_{Sn}, v_{Au}$ are the atomic volumes of Sn and Au atoms. Neglecting the effect of the imaginary part of the atomic scattering amplitude, the scattering can be described in terms of the forward atomic-scattering factor of the atoms. For Au this is given by,

$Z'_{Au}(E) = Z_{Au} + f'_{Au}(E)$ [22] and $\rho_{Au}^{rel} = \dfrac{n_{Au} Z'_{Au}(E)}{\rho_\infty}$ where $n_{Au}$ is the number density of Au atoms and $f'_{Au}(E)$ is the dispersive correction to the real part of the atomic scattering amplitude. In principle, a similar expression describes the effective relative electron density for Sn; however, the present experiment is done near the Au L2 edge for which the dispersion correction to $Z'_{Sn}(E)$ is negligible.

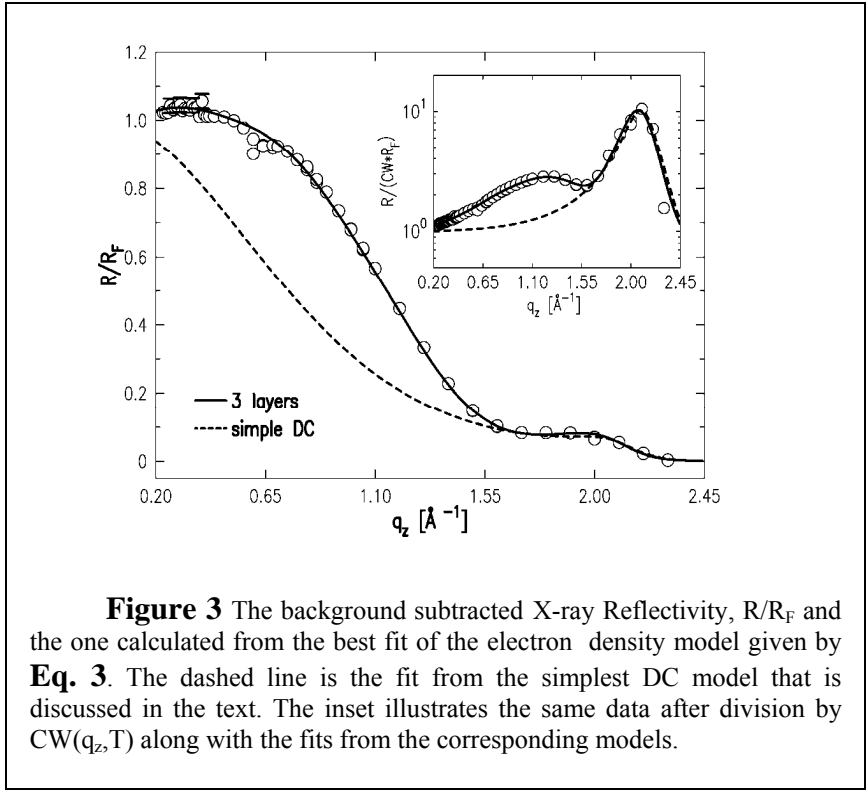

**Figure 3** The background subtracted X-ray Reflectivity, $R/R_F$ and the one calculated from the best fit of the electron density model given by **Eq. 3**. The dashed line is the fit from the simplest DC model that is discussed in the text. The inset illustrates the same data after division by $CW(q_z,T)$ along with the fits from the corresponding models.

The background subtracted Fresnel-normalized reflectivity, $R(q_z)/R_F(q_z)$, at the X-ray energy of 13.15keV is shown in Figure 3. The inset shows the surface structure factor, $|\Phi(q_z)|^2$, which is the Fresnel-normalized data, further normalized by division with the background-subtracted Debye-Waller term. The measured $R(q_z)/R_F(q_z)$ in the $q_z < 1.6\text{Å}^{-1}$ region, is strongly enhanced over the prediction from the DC model (dashed line). The simple DC model for a layered interface with uniform composition that has been used for Ga[18], In[16], and K[17], predicts the surface-induced layering peak at $q_z \sim 2.0$ Å$^{-1}$ but can not produce the broad hump centered at $q_z \sim 1.1$Å$^{-1}$. This feature is an unambiguous indication that either the structure or chemical composition of the near surface region deviates from the DC model.

The best representation of the measured reflectivity in Figure 3 is obtained by numerical fitting of the surface structure factor to the density model given by real space density $\langle \rho(z)/\rho_\infty \rangle$ in Eq. 3. As was mentioned above, the simple DC model that has the same integrated electron density for each layer cannot explain the behavior of the surface structure factor. The electron density of the top few layers had to be modified to fit the data. In many respects this is similar to the case of BiSn liquid alloy that was studied earlier.[11] However, even after allowing for the deviation of the electron density and consequently the composition of the top three layers, the data could not be fit satisfactorily. The acceptable fits shown in Figure 3 were only obtained after allowing the amplitudes, widths and the spacing of these three layers to be modified. As expected from the Gibbs theory the top layer should be rich in the low surface-tension component, Sn and the model obtains an integrated electron density in the first layer that is definitely lower than for any comparable region in the bulk. This corresponds to a Sn rich first layer. The best fits for the second layer also reveals an unambiguously Au-rich region. (Figure 4, Table 1). One word of caution about the meaning of the values in Table 1 is that since the widths of the individual Gaussians are comparable to their spacings the concentrations $x_1, x_2$ and $x_3$ should not be interpreted literally as the concentrations in well defined layers and the electron density profiles shown in Figure 4 are the only realistic representation of the atomic distributions. Nevertheless, the fits do give some information on the atomic concentrations in the third layer. The best fit shown in Figure 4 shows a peak density in the region of the third layer that is less than that of the fourth layer and an average electron density that is less than the bulk value. On the other hand, the Sn to Au fraction at the peak of the electron density at the position of the third layer does correspond to the value of $x_3$ in Table 1 This value is less than the concentration for the bulk alloy, implying that this layer is slightly enriched in Au. In view of the fact that the bulk alloy, $Au_{71}Sn_{29}$, has slightly more than twice as much Au than Sn it may not be surprising to find some Au enrichment in the layer following monolayers of Sn and Au.

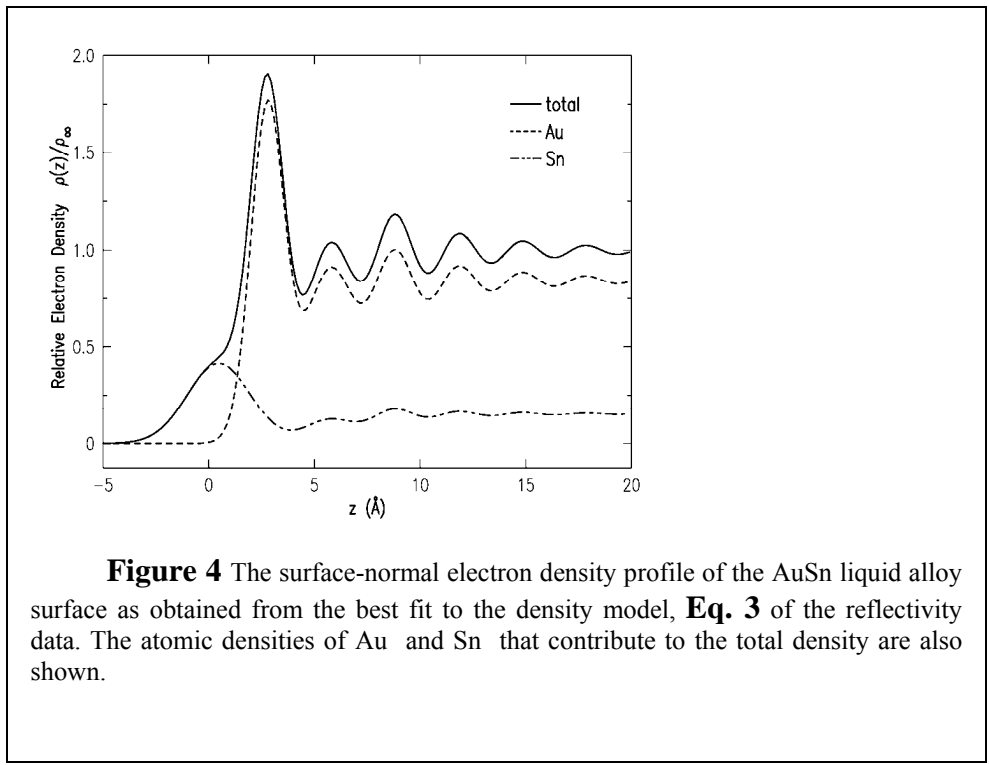

**Figure 4** The surface-normal electron density profile of the AuSn liquid alloy surface as obtained from the best fit to the density model, **Eq. 3** of the reflectivity data. The atomic densities of Au and Sn that contribute to the total density are also shown.

| Sn-concentrations | Positions and spacings (Å) of top layer and others below | Roughness (Å) |
|---|---|---|
| $x_1 = 0.958$ (0.957, -)† | $z_1 = 0.47 \pm 0.01$ | $\sigma_1 = 1.45$ |
| $x_2 = 0.006$ (--, 0.007)† | $z_2 = 2.78 \pm 0.01$ | $\sigma_2 = 0.764$ |
| $x_3 = 0.239$ (0.0, 0.29) | $z_3 = 5.75 \pm 0.01$ | $\sigma_3 = 1.161$ |
| n>3 | $d_0 = 2.93$ | $\sigma_0 = 0.525$  $\bar{\sigma} = 0.514$ |

**Table 1** The best fit parameters from curve-fitting the reflectivity data to the modified distorted crystal model discussed in the text. The Sn-concentrations ($x_1, x_2, x_3$) of the top three modified layers that deviate from the bulk concentration along with the confidence limits, the positions of these layers ($z_1, z_2, z_3$) and their interfacial roughnesses ($\sigma_1, \sigma_2, \sigma_3$) are given. The layer spacing ($d_0$) for layers beyond n=3 is also given along with the roughness parameters that determine the roughness of these layers, $\sigma_0$ and $\bar{\sigma}$.

†The upper and lower confidence limits of $x_1$ and $x_2$, respectively, lie outside the range of the allowed values $x_1$ and $x_2$ can take, namely, 0 to 1.

## 3. SURFACE-SEGREGATION PROFILES AND STATISTICAL THERMODYNAMICS CALCULATIONS

It has been well known for many years that surface-segregation in binary liquid alloys occurs due to the different surface tensions of the constituent atoms in their respective liquid state. In general the surface energy of the alloy is reduced by surface adsorption of the element with the lower surface tension.[1] However, it is only recently that X-ray measurements have yielded the type of data, as shown here, that allow quantitative comparison between theoretical models for the details in how the constituent atoms distribute themselves over the different near surface layers. For example, the surface profiles of BiIn[10] and BiSn[11] have been quantitatively accounted for by calculations based on the solution theories of Guggenheim[24] and its extension by Defay and Prigogine[25] and also by Strohl and King.[26]. These theories assume a lattice model and treat the binary liquid as either an ideal or a regular solution to account for the composition of the surface-segregated layers using only the surface-tensions of the pure components, their atomic radii, and the interaction energy of the components in the case of regular solutions. In contrast for both the BiIn and BiSn systems the near surface layers are relatively well defined and representation of the surface layering with a lattice model is probably better than for the AuSn alloy. Nevertheless, it is interesting to examine the predictions of these models for the current alloy.

Guggenheim assumed a regular solution model and, taking a statistical thermodynamic approach, developed Eq. 4 to predict the surface segregation in the top monolayer in terms of the surface concentration and in interaction energy $w$. The surface tension of the alloy in Guggenheim's theory can be represented as

**Eq. 4**
$$\gamma_{Sn-Au} = \gamma_{Sn} + \frac{kT}{A_{Sn}} \ln\left(\frac{x_1}{x_\infty}\right) + \frac{w}{A_{Sn}}\left[l(1-x_1)^2 - (l+m)(1-x_\infty)^2\right]$$
$$= \gamma_{Au} + \frac{kT}{A_{Au}} \ln\left(\frac{1-x_1}{1-x_\infty}\right) + \frac{w}{A_{Au}}\left[lx_1^2 - (l+m)x_\infty^2\right]$$

where $x_1$ $(1-x_1)$ is the surface concentration in the uppermost layer of component Sn (Au), and the corresponding concentrations in the bulk are $x_\infty$ $(1-x_\infty)$. The surface tensions $\gamma_{Sn}$ and $\gamma_{Au}$ correspond to those of pure liquid Sn and Au extrapolated to the measurement temperature of X-ray reflectivity, T=568K, $\gamma_{Sn}$ = 559 mN/m and $\gamma_{Au}$ = 1258 mN/m.[27] The interaction energy, $w$ is given by, $w=2E_{AB} - (E_{AA}+E_{BB})$ where the $E_{AA}$, $E_{BB}$ and $E_{AB}$ are the different types of atomic interaction energies; however, since the interaction energies are not well known, $w$ is taken to be an adjustable fitting parameter. The atomic areas, $A_{Sn}$ and $A_{Au}$ have been calculated from the atomic radii of the two components, $a_{Sn}=1.41$Å and $a_{Au}=1.44$Å[28] and $l$ and $m$ are the (fractional, $l+2m=1$) lateral and vertical coordination numbers. Their values were assumed to be 0.5 and 0.25 that correspond to those of a hexagonally close-packed lattice with 12 nearest neighbors. With the atomic radii, the hexagonally close-packed coordination numbers, the temperature and the bulk alloy composition x known, the only unknown parameters in Eq.4 are $x_1$, $w$ and $\gamma_{Sn-Au}$. Equating the left and right hand sides, leaves only $x_1$ and $w$ unknown. In principle, one could solve either for $x_1$ for a given value of $w$, or solve for $w$ for a given value of $x_1$ to obtain a predicted surface tension.

Neglecting interactions (ie. $w=0$) and assuming an ideal solution behavior for the AuSn liquid alloy, the calculated concentration of the top most layer turns out to be 0.992 Sn, that is almost a pure monolayer of Sn. While this Sn concentration is close to the experimental value of 0.958 obtained from the curve-fitting analysis of the X-ray reflectivity data, Eq. 4 yields a surface tension of 713 mN/m that is 16% larger than the experimental value of 615 mN/m. On the other hand, the experimental observation of non-zero enthalpy of mixing for the Au-Sn alloys[14] suggests that the interactions between atoms cannot be neglected. Since we are interested in understanding whether the observed composition of the surface layer $x_1$ can explain the surface tension we fixed $x_1=0.958$ and solved Eq. 4 for the interaction energy, $w = -2.20kT$. (The value changes to $w=-2.23kT$ if $x_1$ is set at the lower confidence limit 0.957). Unfortunately, the surface tension obtained with finite $w$ yields an even larger value, 876 mN/m. These results are summarized in the second and third columns of **Table 2**

| Parameters | Guggenheim | | D-P | Experimental Values |
|---|---|---|---|---|
| $w/kT$ | 0 | | -2.20 | |
| $x_1$ | 0.9908 | 0.958* | 0.958* | 0.958 (0.957, --) |
| $x_2$ | | -- | 0.195 (0.194) | 0.006 (--, 0.007) |
| $x_3$ | | -- | 0.305 (0.305) | 0.239 (0.0, 0.29) |
| $x_4$ | | | 0.287 (0.288) | 0.29 |
| $x_\infty$ | | | | 0.29 |
| $\gamma$ (mN/m) | 713 | 815 | 876 | 615 (600,640) |

*Experimental Value

**Table 2** The concentrations and surface tension calculated from the Guggenheim and Defay-Prigogine(DP) as discussed in the text. The interaction parameter $w/kT= -2.20$ that was derived from the Gugenheim theory was used for the D-P theory along with the experimental value for $x_1$. The Sn-concentrations in the brackets correspond to the interaction energy $w/kT=-2.23$ calculated from the lower confidence limit of $x_1=0.957$.

In view of the fact that Guggenheim's theory does not include effects associated with concentration variations beyond the top layer we considered the possibility that the discrepancy in the surface tension might be resolved by inclusion of effects associated with the observed deviation of the concentration in the second and third layers from the bulk values (Figure 4). The Defay-Prigogine (D-P) modification of the Guggenheim theory has been applied successfully to explain the surface-segregation in a BiSn liquid alloy[11, 25] and, in spite of the already mentioned fact that the layers are not as well defined as for the BiSn system we examined the D-P prediction. In this model the Sn-concentration, $x_2$, of the second layer can be calculated from that of the first layer, $x_1$, using Eq. 5 [11,25].

**Eq. 5** $$\ln\left(\frac{(1-x_2)x_\infty}{(1-x_\infty)x_2}\right) - \frac{2\alpha}{kT}\left((1-2m)(1-x_2)-(1-x_\infty)\right) = \frac{2\alpha m}{kT}\left(2-x_\infty+x_1\right)$$

The linearized D-P prediction for the correction to the surface tension is given by

**Eq. 6** $$\gamma_{D-P} = \frac{2mw}{\bar{A}}(1-x_\infty)(x_\infty-x_1) + \gamma_{Gugg}$$

In this modification, where $\bar{A}$ is the average atomic area for $Au_{71}Sn_{29}$, the interaction parameter and/or the value of $x_1$ is unchanged from values obtained from the Guggenheim theory, Eq. 4. Although the Sn-concentrations in the third and fourth layers can be calculated from the concentration in the second and third layers, respectively the surface tension correction to the linearized D-P theory only involves the concentration in the first layer, $x_1$. The results from the D-P equations with the Guggenheim values of $w/kT=-2.20$ and $x_1=0.958$ are shown in the fourth column of **Table 2**.

　　The agreement with experiment is at best only qualitative. The first problem is, as already explained, the association between the concentrations in the layered models of Guggenheim and Defay-Prigogine and with the Gaussian amplitudes of the distorted crystal model is not strict. It is encouraging that both models predict surface layers that are essentially pure Sn, in agreement with the experimental best fit. On the other hand, it is hard to know how one should interpret the discrepancy between the nearly zero concentration of the Sn in the second layer of the distorted crystal model and the 20 atm% Sn concentration in the second layer of the D-P model. For example, the tail of the Gaussian peak of distorted crystal model that represents the first layer of Sn partially overlaps the Gaussian peak that represents the second layer that is Au. For these Gaussians the ratio of the Sn to Au density at the peak position of the Au Gaussian, $z=2.78$, corresponds to a Sn concentration of nearly 30% that is not so different from the 20 % of the D-P model. On the other hand there seems to be a clear disagreement between the D-P prediction for a third layer is slightly enriched in Sn and the result from the distorted crystal model in which the third layer is enhanced in Au.

Strohl-King (S-K) have developed an alternative to the D-P theory which unfortunately has the same shortcoming of assuming well defined layers.[26] This theory can be used to estimate the concentrations in the various layers by an iterative procedure in which the concentrations in the various layers can be adjusted to match the measured concentrations by varying the interaction parameter. The basic equation for the concentrations is

**Eq.7**

$$1-x_n = (1-x_\infty)\exp\left(\frac{w}{kT}\left\{x_\infty^2 - \left(\frac{lx_n^2}{l+m}\right) + \left(\frac{m}{l+m}\right)\left[(x_{n-1})^2 + (x_{n+1})^2\right]\right\} + \frac{A_{Au}(\gamma_{Sn}-\gamma_{Au})x_n}{kT}\right)$$

The calculated values that are shown in the 2$^{nd}$ column of (Table 3) correspond to the value of the interaction parameter $w/kT$ that is obtained by minimization of $\chi^2 \equiv \sum_{n=1}^{10}(x_n - x_n^{\exp})^2$. The optimum interaction energy $w=-1.78\ kT$ is about 20% lower than the D-P value of -2.2kT. Although the S-K theory does predict a near vanishing of the Sn concentration in the the second layer it is hard to know the significance of this. For completeness the values in the 3$^{rd}$ column of Table 3 demonstrate that the results from D-P for $w/kT=-1.78$ are not significantly different from the D-P values shown in **Table 2** for $w/kT=-2.2$ that were obtained from the Guggenheim equation (Eq. 4) when $x_1$ was fixed at the experimental concentration.

| Parameters | S-K | D-P(*) | Experimental |
|---|---|---|---|
| $x_1$ | 0.995 | 0.969 | 0.958 (0.957, --) |
| $x_2$ | 0.009 | 0.206 | 0.006 (--, 0.007) |
| $x_3$ | 0.309 | 0.301 | 0.239 (0.0, 0.29) |
| $x_4$ | 0.287 | 0.288 | 0.29 |
| $\gamma$(mN/m) | -- | 847 | 615 |

**Table 3** The Sn concentrations calculated from S-K by optimization of w/kT=-1.78. The third column lists the values calculated from D-P with the interaction parameter that was obtained from S-K theory.

Various thermophysical properties like the enthalpy of mixing, surface-tension, viscosity and the surface-compositions of the liquid AuSn alloy have been calculated recently by Novakovic and coworkers[20] using what they refer to as a 'complex formation model' that is a combination of Guggenheim and other statistical-thermodynamic theories. Although the surface-concentration of Sn they obtain (~ 0.65) does exceed the bulk value it is still significantly smaller than the value extracted from the X-ray reflectivity data. Similarly, for 29 atm % Sn their surface tension (~960 mN/m) is also much larger than the X-ray determined value of 615 mN/m. On the other hand, the interaction energy parameters, used to calculate the surface concentration profiles, -2.42 (at 550°C) and -1.25 (at 1100°C) extrapolate linearly to -3.0 at 300°C, that is only ~25% larger than the value of -2.2 obtained from Guggenheim and D-P theories. This can be compared to bulk interaction energy $w_{bulk} = \Delta H_m / x(1-x)$ =-9.98kT that is calculated from the enthalpy of mixing, $\Delta H_m$=-9.66 kJ/mole, which is nearly a factor of four larger than the surface interaction energy obtained here.[14] In fact, a similar disagreement was observed for SnBi[11], where the argument was made that that bulk thermodynamic quantities often yield inaccurate values for surface quantities.

## 4. CONCLUSIONS

X-ray reflectivity studies show that the atomic-level surface-segregation occurs in the eutectic AuSn liquid alloy and extends up to three layers. The top layer is almost a pure monolayer of the low surface-tension component, Sn. The extent of segregation observed is similar to that found in BiSn alloys where the surface-segregation probably extends up to three layers with the top most layer very rich in the low-surface tension component, Bi. The $Au_{71}Sn_{29}$ system differs from the nearly equimolar $Bi_{43}Sn_{57}$ in that for the latter the enrichment by layers was Bi:Sn:Bi while here it appears to be Sn:Au:Au. The extent of segregation observed is more pronounced than in BiIn that also has negative enthalpy of mixing. All of the liquid alloys studied so far by X-ray reflectivity show the segregation of the low surface-tension component to the surface independent of whether their enthalpy of mixing in the bulk is positive or negative. This behavior shows that the surface energy plays a dominant role in surface-segregation. While the simple theories of surface-segregation can qualitatively explain the surface-segregation profiles reasonably well, they cannot account for the measured surface tension and the agreement regarding concentrations is not quantitative. For the AuSn liquid alloy studied here the existing theories cannot satisfactorily explain both the surface-segregation profiles and the surface tension together. There clearly is a need for improvements in the theory of the surfaces of liquid metal alloys.

Finally we should point out that equally good fits to the reflectivity can be obtained by a different model in which the top surface layer is Au-rich rather than Sn rich. This is not completely surprising in that it is known that the lack of phase information due to the fact that the reflectivity depends on the absolute value of the surface structure factor implies an ambiguity in interpretation.[29] Fortunately, the physical constraint imposed by the Gibbs adsorption allows the Au-rich surface model to be dismissed. Aside from this the Guggenheim, and Defay-Prigogine theories would require an interaction energy parameter of 72 ($w/$kT) in order to explain the Au rich surface composition model. Even after allowing for the above mentioned uncertainty relating bulk and surface quantities, this very large positive interaction parameter would imply a very large positive enthalpy of mixing that is not observed in the bulk and is certainly not physically realistic.

## 5. ACKNOWLEDGEMENTS

This work has been supported by the U.S. Department of Energy through the grant DE-FG02-88-ER45379, ChemMatCARS is principally supported by the National Science Foundation/Department of Energy under the grant CHE0087817. The Advanced Photon Source is supported by the U.S. Department of Energy, Basic Energy Sciences, Office of Science, under Contract No: W-31-109-Eng-38.